\documentclass[%
reprint,
amsmath,amssymb,
aps,
]{revtex4-2}
\usepackage{graphicx}
\usepackage{dcolumn}
\usepackage{bm}
\usepackage[colorlinks,linkcolor=black,urlcolor=black,citecolor=blue]{hyperref}

\begin{document}
	
\preprint{APS/123-QED}

\title{Controllable optical bistability with perfect photon absorption}
\author{Miaodi Guo}%
\email{guomiaodi@xatu.edu.cn}
\affiliation{School of Sciences, Xi'an Technological University, Xi'an 710021, China
}
\date{\today}

\begin{abstract}
We propose a scheme for controlling nonlinear coherent perfect absorption (CPA) in a three-level $\Lambda$-type atom-cavity system. Generally, the normally nonlinear CPA and the bistable CPA can be attained at the different frequencies of an input probe field. With a coherent control field coupling one of the ground states and the excited state of the atoms, two types of CPA can be attained at the same frequency. Besides, for the bistable CPA, the highest output-input ratio of the high stable state and the low stable state is approximately 1, and the bistable region is controllable with the control field. The controllable bistable CPA may have potential  applications in optical bistable switching and optical logic devices.
\end{abstract}

\maketitle

\section{introduction}
As a typical nonlinear phenomenon, the optical bistability (OB) has played an important role in the field of all-optical processing, such as, all-optical switching, optical memories, and logical gates \cite{PRL108/263905,Sheng13,Nozaki2012,DelBino:21}. The nonlinearity can be observed in the nonlinear medium within an optical resonator, e.g., atom-cavity system \cite{Sawant2016a,Megyeri2018,Abdelaziz2020} and fiber ring cavities \cite{Li2017}, or in nonlinear directional couplers \cite{Ali2016}, nanostructures \cite{Nozaki:15,Jiang2020} and graphene plasmonics \cite{Li:21}, etc. Recently, the OB is realized with coherent perfect absorption (CPA) in a two-level atom-cavity system \cite{Agarwal2016,Xiong2020}.  CPA \cite{Zhang2012b,Jeffers2019}  as the realization of extreme absorption arises from the interference of two counterpropagating light fields, which is related to the lasing process by time reversal \cite{Wan2011} and can also be achieved in solid-state Fabry-Perot devices and metamaterials \cite{Horng2020,Kang2018}. Besides the manipulation on photon absorption, the operation point of CPA can be tuned to be in the linear or the bistable domain, which can potentially be applied in some optical devices \cite{Suwunnarat2019}. And when CPA is at the low stable branch of a bistable switching, it is possible to increase the output-input ratio between the high stable value and the low stable value, which can be applied to increase the switching efficiency \cite{Wang2017a}.

Generally, the CPA occurs only within the certain frequency range when the atom-cavity system is determined \cite{Agarwal2016,Xiong2020,Wei2018}. For example, in the two-level atom-cavity system \cite{Agarwal2016}, dual-frequency CPA modes at $\Delta_p=\pm4.5\Gamma$ are attained for a given input intensity $|a_{in}|^2\approx55$ ($\Delta_p$ is the frequency detuning of the probe field, and $\Gamma$ is the decay rate of atomic excited level). The frequency of CPA is variable only with some certain intensities of the probe fields. However, in a three- or four-level atom-cavity system \cite{Wang2017a,Wu2016}, with the quantum coherence and interference between two transitions of the atoms, the frequency range of CPA can be controlled, and the linear and nonlinear CPA can be observed at the same frequency of a probe field \cite{Guo2021}. 

\section{Theoretical analysis}
\begin{figure}[!hbt]
	\centering
	\includegraphics[width=5cm]{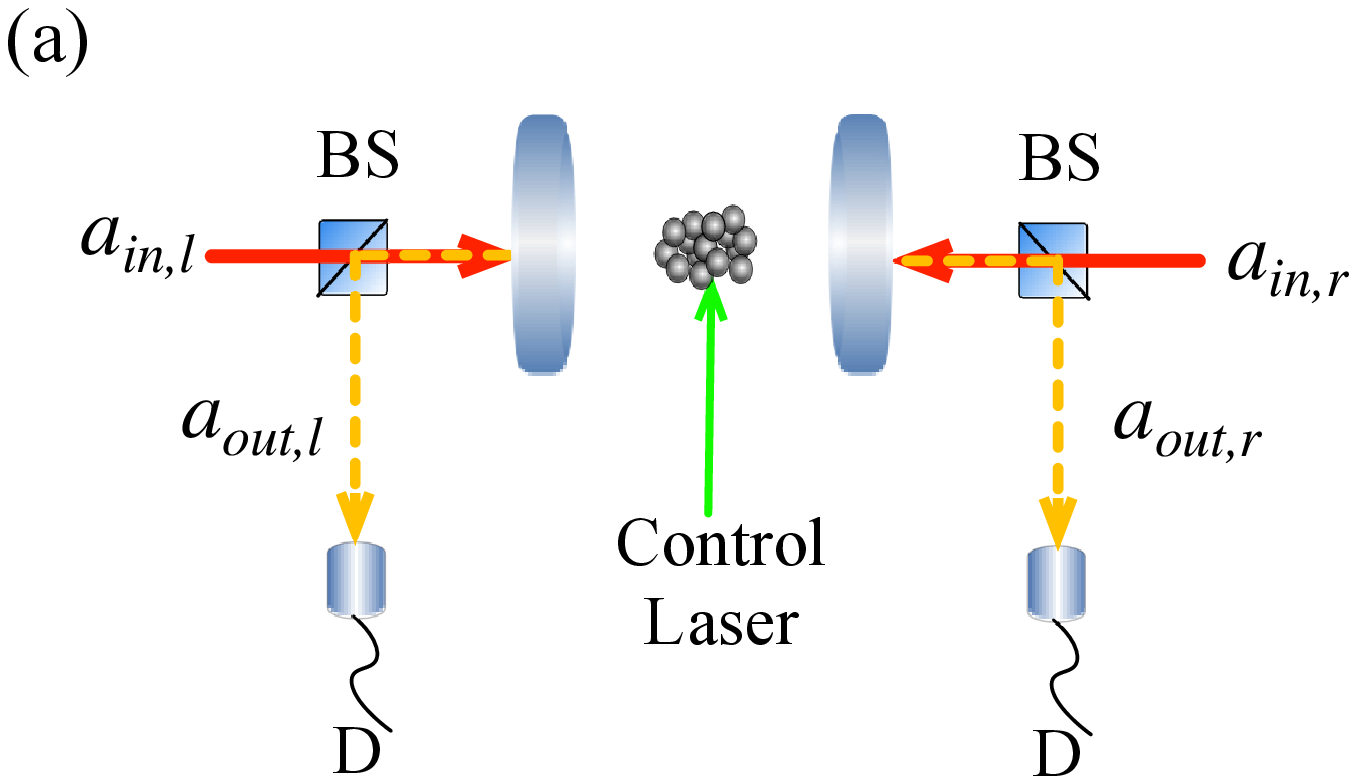}
	\includegraphics[width=5cm]{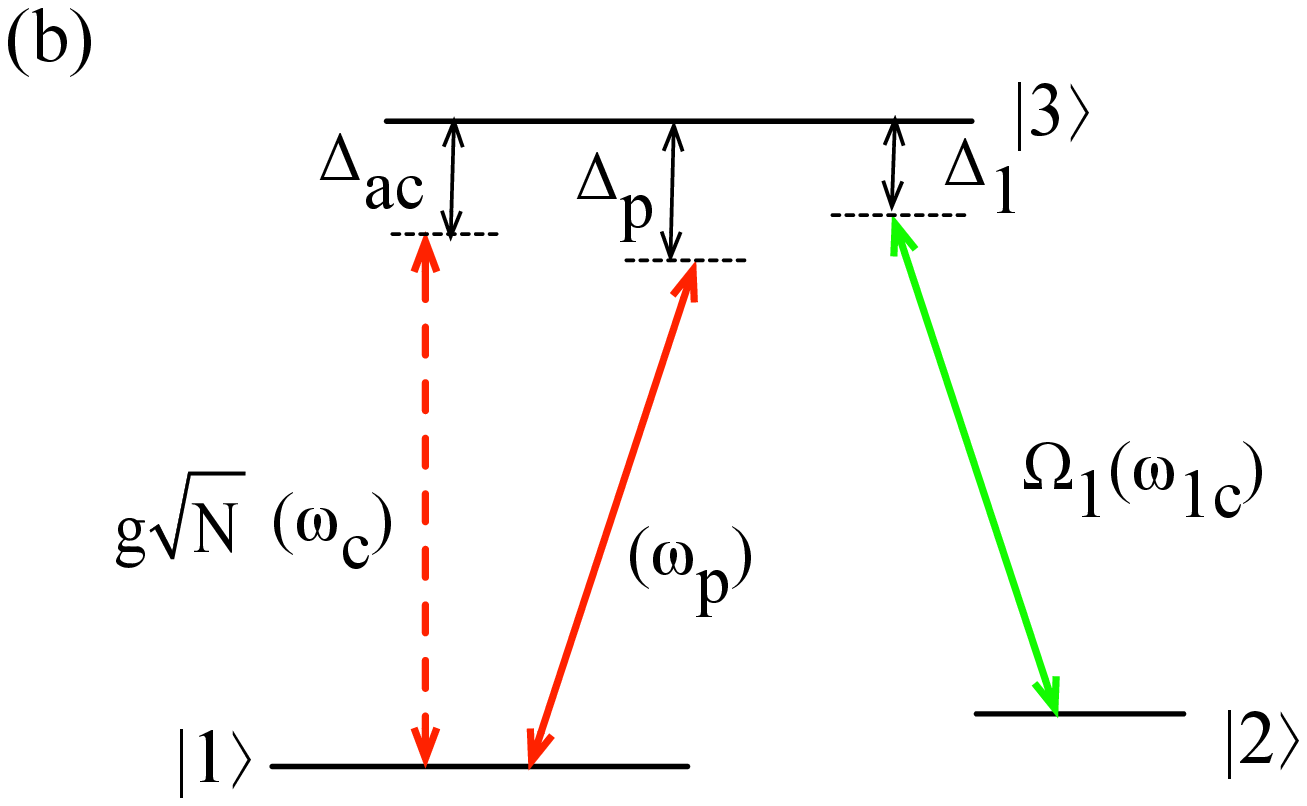}
	\caption{(a) Schematic diagram of a two-sided cavity with some (b) three-level atoms.}\label{fig-system}
\end{figure}
In this Letter, we analyze one kind of unconventional OB with CPA in a three-level $\Lambda$-type atom-cavity system. With a relative phase and a coherent control field, the bistable CPA is tunable and controllable. The schematic diagram of the system is shown in Fig. \ref{fig-system}. We consider a system consisting of some cold three-level atoms trapped in a two-sided optical cavity. Two coherent probe lasers $a_{in,l}$ and $a_{in,r}$ are injected into the cavity through two beam splitters (BS), and drive the atomic transition $|1\rangle\rightarrow|3\rangle$ with a frequency detuning $\Delta_p=\omega_p-\omega_{31}$. $\Delta_{ac}=\omega_c-\omega_{31}$ is the frequency detuning between the cavity mode and the atomic transition $|1\rangle\rightarrow|3\rangle$. A control laser is coupled to the atomic transition $|2\rangle\rightarrow|3\rangle$ with a frequency detuning $\Delta_1=\omega_{1c}-\omega_{32}$. Two detectors (D) are applied to measure the output fields $a_{out,l}$ and $a_{out,r}$. Here $\omega_c$ is the frequency of the cavity mode, $\omega_p$ and $\omega_{1c}$ are frequencies of the probe and control lasers, and $\omega_{mn}$ ($m,n=1,2,3$) is the frequency of corresponding atomic transition. For accessible implementation, the hyperfine energy levels 5$S_{1/2}$ $F=1$, 5$S_{1/2}$ $F=2$ and 5$P_{3/2}$ of $^{87}$Rb can be selected as levels $|1\rangle$, $|2\rangle$, and $|3\rangle$, respectively.

The Hamiltonian of the system is \cite{Wang2017a},
\begin{equation}
\begin{split}
		H=&-\hbar  \sum _{j=1}^N [(\Delta_p-\Delta_1) \sigma_{22}^j+\Delta_p \sigma_{33}^j]-\hbar  (\Delta_p-\Delta_{ac})a^{\dagger}a\\&-\hbar  \sum _{j=1}^N (g a \sigma_{31}^j+\Omega_1 \sigma_{32}^j)+H.C.\label{equ-H}
\end{split}
\end{equation}
where $\hbar$ is the reduced Planck constant, $N$ is the number of the atoms inside the cavity, $\sigma_{mn}^{j}=|m\rangle\langle n|$ ($m,n=1,2,3$) is the atomic operator, $a^{\dag}$($a$) is the creation (annihilation) operator of cavity photons, $g=\mu_{13}\sqrt{\omega_{c}/(2\hbar\varepsilon_{0}V)}$ is the cavity-QED coupling coefficient ($\varepsilon_{0}$ is the free space permittivity and $V$ is the cavity mode volume), $\Omega_1=\mu_{23} E/\hbar$ is the Rabi frequency of the control laser ($E$ is the field amplitude and $\mu_{mn}$ is the matrix element of the electric dipole moment), and  $H.C.$ denotes the Hermitian conjugate.

In semiclassical approximation, we treat the expectation values of field operators as the corresponding fields, e.g., $\langle a\rangle=\alpha$ and $\langle a^{\dagger}\rangle=\alpha^*$ \cite{Sawant2016a}. According to the following differential equations
\cite{ScullyMOZubairyMS2001},
\begin{equation}
\begin{split}
	\dot{\rho}&=\frac{1}{i\hbar}[H,\rho]-\frac{1}{2}\{\gamma,\rho\},\\
	\dot{a}&=\frac{1}{i\hbar}[a,H]-\frac{\left(\kappa _l+\kappa _r\right)}{2}a+\sqrt{\frac{\kappa _l}{\tau} } a_{in,l}+\sqrt{\frac{\kappa _r}{\tau}} a_{in,r},\label{equ-dot}
	\end{split}
	\end{equation}
we have,
\begin{widetext}
\begin{equation}
		\begin{split}
		&\dot{\rho_{11}}=\frac{\Gamma }{2} \rho_{33}+i g (a^\dagger \rho_{13}-a \rho_{31}),\\
		&\dot{\rho_{12}}=[i (\Delta_p-\Delta_1)-\gamma_{12}]\rho_{12}-i g a  \rho_{32}+i \Omega_1 \rho_{13} ,\\
		&\dot{\rho_{13}}=(i \Delta_p-\frac{\Gamma}{2})\rho_{13}+i g a (\rho_{11}-\rho_{33})+i \Omega_1\rho_{12} ,\\
		&\dot{\rho_{22}}=\frac{\Gamma}{2} \rho_{33}+i \Omega_1 (\rho_{23}-\rho_{32}),\\
		&\dot{\rho_{23}}=(i \Delta_1-\frac{\Gamma }{2})\rho_{23}+i g a  \rho_{21}+i \Omega_1 (\rho_{22}-\rho_{33}),\\
		&\dot{\rho_{33}}=-\Gamma \rho_{33}+i g (a  \rho_{31}-a^\dagger \rho_{13})+i \Omega_1 (\rho_{32}-\rho_{23}),\\
		&\dot{a}=i \left(\Delta _p-\Delta _{ac}\right)a+i g N \rho _{13}-\frac{\left(\kappa _l\!+\!\kappa _r\right)}{2}a+\sqrt{\frac{\kappa_l}{\tau}} a_{in,l}+\sqrt{\frac{\kappa _r}{\tau}} a_{in,r},\label{equ-dots}
	\end{split}
	\end{equation}	
\end{widetext}
where $\langle n|\gamma|m\rangle=\Gamma_{n}\delta_{nm}$, $\Gamma$ is the decay rate of atomic level $|3\rangle$, $\gamma_{12}$ is the decoherence rate between atomic levels $|1\rangle$ and $|2\rangle$, $\kappa_l$ ($\kappa_r$) is the field decay rate from left (right) cavity mirror, and $\tau$ is the photon round trip time inside the cavity. 

We consider a symmetric cavity with $\kappa_l=\kappa_r=\kappa$, and assume the control laser is resonant with the atomic transition $|2\rangle\rightarrow|3\rangle$, i.e., $\Delta_1=0$, the intracavity field $\alpha$ under steady state can be written as,
\begin{widetext}
\begin{equation}
\alpha=\frac{\sqrt{\kappa/\tau} \left(\alpha _{in,l}+\alpha _{in,r}\right)}{X-\frac{2  \Omega _1^2 g \alpha \left[2g^2 \left| \alpha \right|^2 \left(A-2 \gamma _{12} \Delta _p\right)+A\left(\Gamma\gamma _{12} +i \Gamma  \Delta _p+2 i \gamma _{12} \Delta _p-2 \Delta _p^2+2 \Omega _1^2\right)\right]}{\Gamma  \Omega _1^2 \left\{B+4\left[\gamma _{12}^2 \Delta _p^2+\left(\Delta _p^2-\Omega _1^2\right)^2\right]\right\}+g^2 \left| \alpha \right|^2 \left[C+8 \gamma _{12} \Omega _1^2 \left(\Delta _p^2+6 \Omega _1^2\right)\right]}},\label{equ-a}
\end{equation}
\end{widetext}
where $X=\kappa-i \left(\Delta _p-\Delta _{ac}\right)$, $A=\Gamma(\Delta_p+i\gamma_{12})$, $B=\Gamma ^2(\gamma _{12}^2+\Delta _p^2)+4  \Gamma \gamma _{12} \Omega _1^2$, and $C=\Gamma ^3 (\gamma _{12}^2+\Delta _p^2)+8 \Gamma ^2 \gamma _{12}  \Omega _1^2+4 \Gamma  \Omega _1^2 (6 \gamma _{12}^2+4 \Delta _p^2+3 \Omega _1^2)$. Although the higher order items of $\alpha$ are neglected, it is clear that Eq. (\ref{equ-a}) is a cubic equation of $\alpha$. Therefore, the output fields are nonlinearly dependent on the input probe fields, and can be attained according to the following input-output relations \cite{Agarwal2016,WallsDFandMilburnGJ2007},
 \begin{equation}
 	\begin{split}
	\langle a_{out,l}\rangle&=\sqrt{\kappa_l \tau}\langle a\rangle-\langle a_{in,l}\rangle,\\
	\langle a_{out,r}\rangle&=\sqrt{\kappa_r \tau}\langle a\rangle-\langle a_{in,r}\rangle.\label{equ-in-out}
	\end{split}
	\end{equation}
We assume $a_{in,l}=|a_{in}|e^{i\varphi}$ and $a_{in,r}=|a_{in}|$, the system can be acted as a perfect absorber when $\langle a_{out,l}\rangle=\langle a_{out,r}\rangle=0$, i.e., $a_{in,l}=a_{in,r}$ and $\varphi=2n\pi$,.

\section{Numerical results}
 \begin{figure}[!hbt]
	\centering
	\includegraphics[width=5cm]{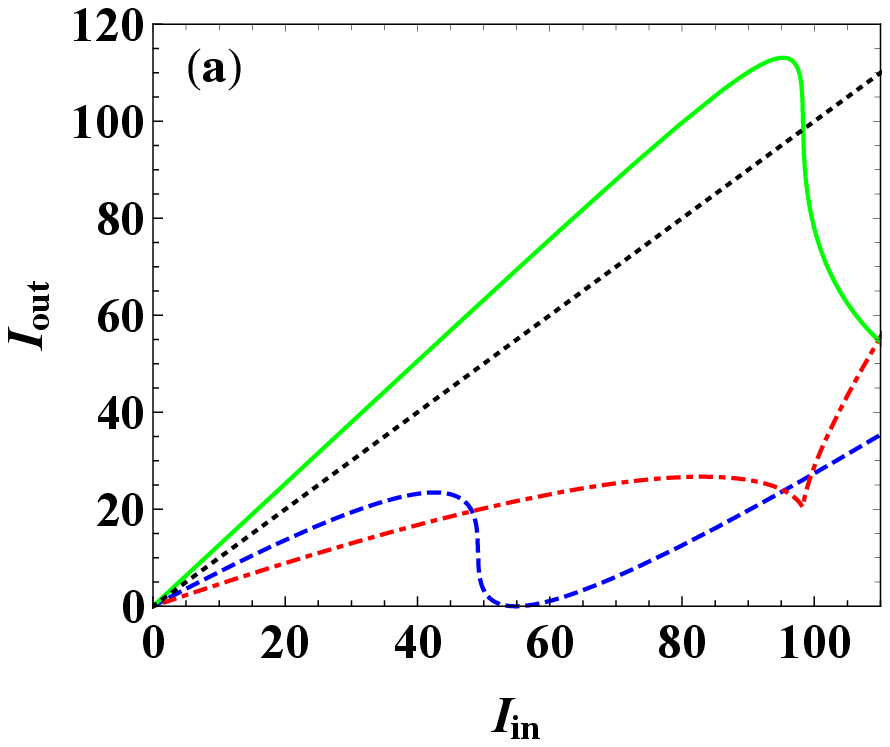}
	\includegraphics[width=5cm]{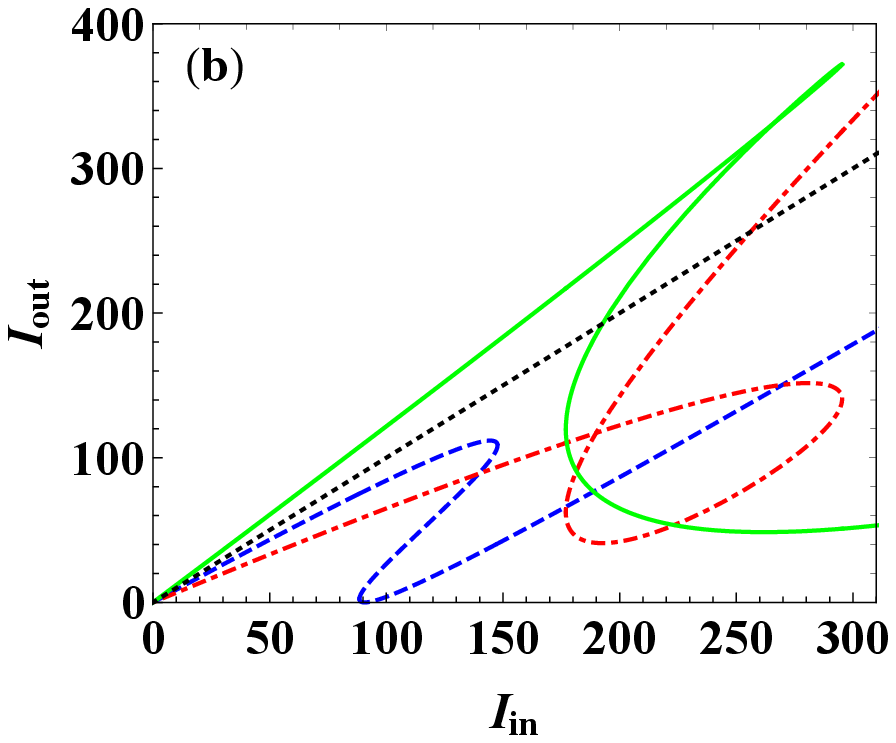}
	\caption{Output intensity versus input intensity for (a) $\Delta_p=6.5\Gamma$, $\Delta_{ac}=-5.5\Gamma$, and (b) $\Delta_p=6\Gamma$, $\Delta_{ac}=-4.5\Gamma$ with $\varphi=0$ (dashed blue line), $\varphi=\pi/2$ (dot-dashed red line for the right output and solid green line for the left output ), and $\varphi=\pi$ (dotted black line). The parameters are $\kappa=\Omega_1=\Gamma$, $g\sqrt{N}=10\Gamma$, $\tau=0.01/\Gamma$, and $\gamma_{12}=0.001\Gamma$. }\label{fig-Iout}
\end{figure}

 When $\varphi=0$, two output fields are identical and CPA occurs with certain input intensity. It is calculated that the frequency range of CPA under these conditions \cite{Guo2021} is $\Delta_p=-7.1\Gamma\sim7.1\Gamma$, when $\Delta_p$ is close to the threshold value, linear CPA can be attained with weak input intensity \cite{Agarwal2016,Guo2021}. When $|\Delta_p|$ decreases from the threshold value, e.g., $\Delta_p=6.5\Gamma$ and $\Delta_p=6\Gamma$, nonlinear CPA can be observed as shown by the dashed blue curves in Fig. \ref{fig-Iout}. When $\Delta_p=6.5\Gamma$, the cavity field $|\alpha|^2$ has one real-value solution in Eq. (\ref{equ-a}), and the system is driven into a normally nonlinear CPA regime as shown in Fig. \ref{fig-Iout}(a). When $\Delta_p=6\Gamma$, the cavity field $|\alpha|^2$ has three real-value solutions, and the system is driven into the unconventional bistable CPA regime as shown in Fig. \ref{fig-Iout}(b). Changing the relative phase, e.g. $\varphi=\pi/2$, two output fields are generally different and the system is out of the bistable CPA domain as shown by the solid green and dot-dashed red curves in Fig. \ref{fig-Iout}. However, more complicated bistable curves are presented with the strong input intensity as shown in Fig. \ref{fig-Iout}(b). When $\varphi=\pi$, two input probe fields interfere destructively and no light can be coupled to the cavity, which causes $I_{out,l}=I_{out,r}=I_{in}$. Therefore, $I_{out}$ is linearly dependent on $I_{in}$ and coherent perfect reflection (CPR) is attained under the condition as shown by the dotted black lines in Fig. \ref{fig-Iout}. By relative phase $\varphi$, the state transfer from CPA to CPR can be realized, which has the potential application in the research of optical switching or phase gates.

\begin{figure}[!hbt]
	\centering
	\includegraphics[width=5cm]{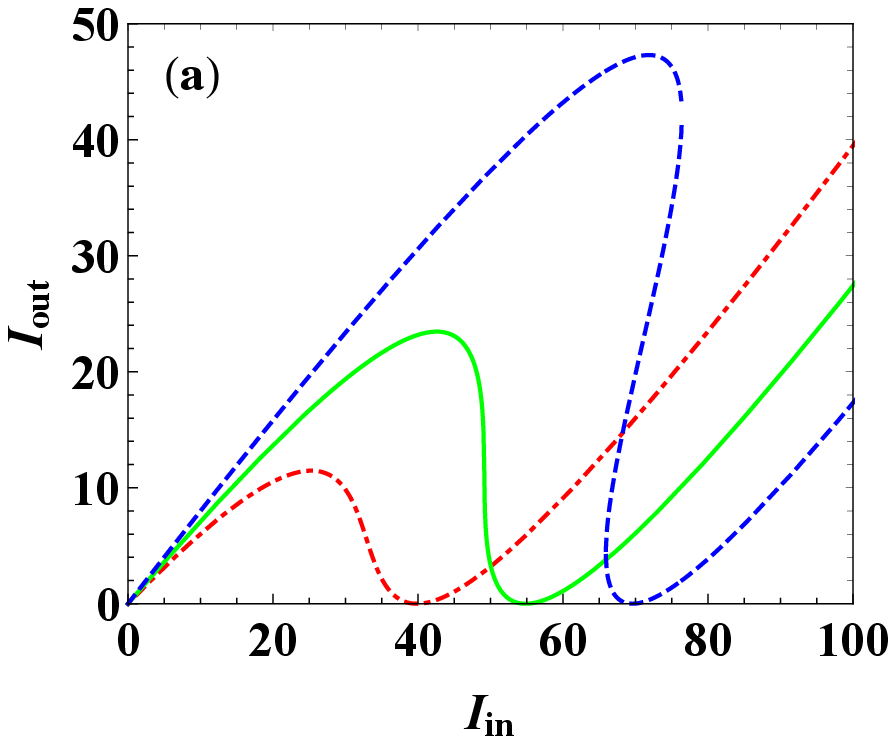}
	\includegraphics[width=5.05cm]{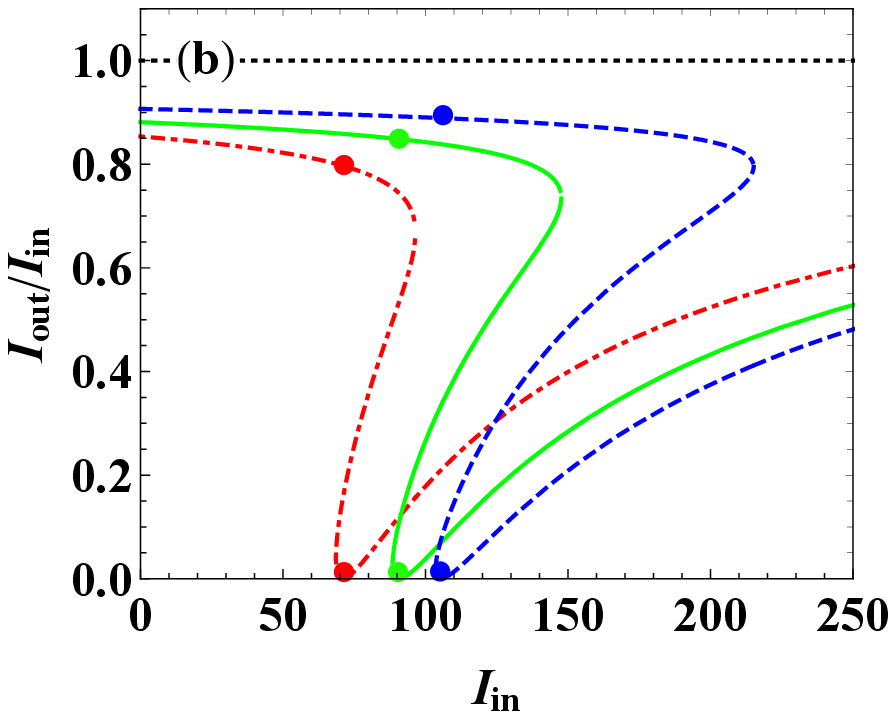}
	\includegraphics[width=5.15cm]{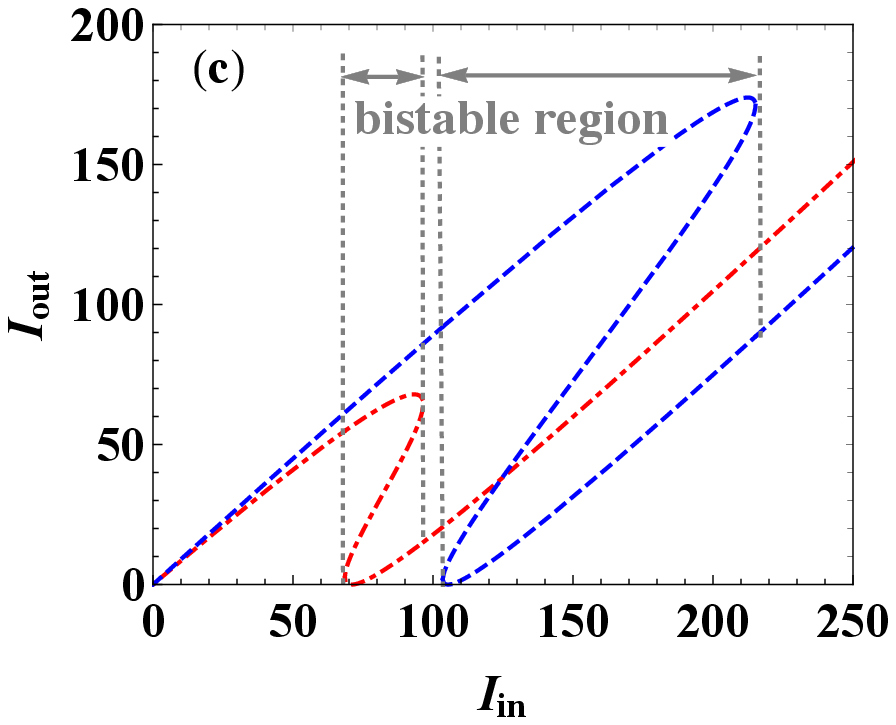}
	\caption{Output intensity (a) (c) and output-input ratio (b) versus input intensity with $\Omega=0.5\Gamma$ (dot-dashed red line), $\Omega=\Gamma$ (solid green line), and $\Omega=1.5\Gamma$ (dashed blue line). The other parameters are the same with those in Fig. \ref{fig-Iout}.}\label{fig-Iout-Iin}
\end{figure}

When $\Omega_1=0$, the scheme can be acted as a two-level CPA system \cite{Agarwal2016}, where the CPA can be realized at some specific frequencies which are not tunable. However, when $\Omega_1\neq0$, the coupling between the atomic levels $|2\rangle$ and $|3\rangle$ forms a destructive quantum interference between two absorption channels of the transition $|1\rangle\rightarrow|3\rangle$. And this will induce an interference control on CPA condition. Therefore, the frequency of CPA is tunable, and the transfer between the coherent non-prefect absorption and the CPA at a certain frequency can be realized \cite{Guo2021} in the three-level CPA system. And the relevant results have been presented in reference \cite{Guo2021}.

In a two-level nonlinear CPA system, the bistable CPA can be observed under certain frequencies \cite{Agarwal2016,Xiong2020}. However, the frequencies can be tunable in the three-level nonlinear CPA system, where the susceptibility is nonlinearly dependent on $\Omega_1$ and strengthening the control laser will improve the nonlinear items of media susceptibility. For example, when $\Omega_1=0.5\Gamma$ and $\Omega_1=\Gamma$, the monostable CPA appears at $\Delta_p=6.5\Gamma$ as shown by the dot-dashed red and solid green curves in Fig. \ref{fig-Iout-Iin}(a). Increasing $\Omega_1$ as  $\Omega_1=1.5\Gamma$, the bistable CPA can be observed at the frequency as shown by the dashed blue curve. Besides, in the bistable domain, the high stable state at the critical point with CPA is close to the CPR when $\Omega_1$ is increased. For example, when $\Delta_p=6\Gamma$, the output-input ratios for the high stable values are 0.79, 0.85, 0.89 corresponding to the low stable values of CPA when $\Omega_1=0.5\Gamma$, $\Omega_1=\Gamma$, $\Omega_1=1.5\Gamma$, respectively, as shown by the red, green and blue points in Fig. \ref{fig-Iout-Iin}(b), where the CPR is represented by the dotted black line. When $\Omega_1=4\Gamma$, it is calculated that the ratio of the high stable value is 0.98, which means an optical-bistability switching can be realized with the CPA state as a low stable state and the near-CPR state as a high stable state. And because of the unconventional bistable property, the bistable switching can be realized under a weak probe field. In addition, when $\Delta_p$ decreases from the threshold value, e.g. $\Delta_p=6\Gamma$, the bistable region becomes wider with the increase of the intensity of the control field. For example, when $\Omega_1=0.5\Gamma$, the bistable domain is observed when the intensity of the input field is approximately from $I_{in}=70$ to $I_{in}=95$ as shown by the dot-dashed red curve in Fig. \ref{fig-Iout-Iin}(c). Increasing $\Omega_1$ as $\Omega_1=1.5\Gamma$, the bistable domain appears with $I_{in}\approx105 \sim I_{in}\approx220$ as shown by the dashed blue curve. With the coherent control field, the tunable and controllable bistable CPA is achieved.

\section{conclusion}
In conclusion, we have analyzed the controllable bistable CPA in a three-level atom-cavity system. With the relative phase of two input probe beams, the output intensity is manipulated in the linear regime or the bistable regime. The CPR and the nonlinear CPA are attained when the relative phase is $\varphi=(2n+1)\pi$ and $\varphi=2n\pi$, respectively. When $\varphi$ is at other values, more complicated nonlinear pattern of the output intensity is attained. With a coherent control field, the bistable CPA is tunable and the bistable region is controllable. In addition, the output-input ratio of the high stable state and the low stable state (CPA) is up to 1, which means the high stable state is close to the CPR. And this may have potential applications in logical elements of optical communication and calculation.

\section*{Acknowledgments}
This work is  supported by the Natural Science Foundation of Shaanxi Provincial Department of Education (Grant No. 20JK0682).

\end{document}